# Ab-initio Electronic and Structural Properties of Rutile Titanium Dioxide

Chinedu, E. Ekuma and Diola Bagayoko

[1]Department of Physics and Astronomy, Louisiana State University, Baton Rouge, LA 70803, U.S.A.

[2]Department of Physics, Southern University and A&M College, Baton Rouge, LA 70813, U.S.A

Ab-initio, self-consistent electronic energy bands of rutile $TiO_2$ are reported within the local density functional approximation (LDA). Our first principle, non-relativistic and ground state calculations employed a local density functional approximation (LDA) potential and the linear combination of atomic orbitals (LCAO). Within the framework of the Bagayoko, Zhao, and Williams (BZW) method, we solved self-consistently both the Kohn-Sham equation and the equation giving the ground state charge density in terms of the wave functions of the occupied states. Our calculated band structure shows that there is significant $O_{2p}$-$Ti_{3d}$ hybridization in the valence bands. These bands are well separated from the conduction bands by an indirect band gap of 2.95 eV, from $\Gamma$ to R. Consequently, this work predicts that rutile $TiO_2$ is an indirect band gap material, as all other gaps from our calculations are larger than 2.95 eV. We found a slightly larger, direct band gap of 3.05 eV, at the $\Gamma$ point, in excellent agreement with experiment. Our calculations reproduced the peaks in the measured conduction and valence bands densities of states, within experimental uncertainties. We also calculated electron effective mass. Our structural optimization led to lattice parameters of 4.65 Å and 2.97 Å for $a_o$ and $c_o$, respectively with a $u$ parameter of 0.3051 and a bulk modulus of 215 GPa.

1.  **Introduction [and Motivation]**

Titania ($TiO_2$) is one of the most studied transition metal oxides. It crystallizes in four distinct polymorphs: rutile, anatase, brookite and an n-$TiO_2$ in order of decreasing abundance.[1,2] Of these forms, rutile $TiO_2$ is the most stable.

Over the past several decades, $TiO_2$ has been extensively studied both experimentally and theoretically. It has interesting, physical and chemical properties[3] that can be harnessed for diverse technological applications. $TiO_2$ is not just a better photocatalyst in heterogeneous photocatalytic applications[4] due to its functionality, but also a promising material for photochemical applications[5]. It has high dielectric constant and an excellent optical transmittance in the visible and near infrared regions. It has been used in UV induced electron photo-excitation,[6,7] as pigment in paint,[8,9] and in hydrogen production.[10] It is used in sensors,[9] transparent conducting oxides,[11] opacifiers (due to its high reflectivity across the visible spectrum),[1] and in photocatalysts for solar energy utilization and environmental clean-up.[12,13,14] $TiO_2$ has also been used in resistive memories.[15] It is commonly used in electronic in thin film capacitors[16] and in the realization of spintronic devices.[9,17,18] It is employed in the fabrication of antireflection coatings, interference filters, optical waveguides,[19] gas sensors,[1] and as a ferroelectric material at low pressures.[20] Due to its non-toxicity, long term stability in aqueous solutions, and its chemical inertness, rutile $TiO_2$ is an important material in aqueous radiation and photochemistry.[8,21]

The numerous technological applications of $TiO_2$ partly motivated extensive, experimental studies of the properties of rutile $TiO_2$ using different techniques such as X - ray photoemission spectroscopy (XPS),[22-26] electron-energy loss spectroscopy (EELS),[27-31] ultra-violet photoemission spectroscopy (UPS),[32] Auger emission spectroscopy (AES), total energy yield spectroscopy,[33] X - ray emission spectroscopy (XES),[34,35] X - ray absorption spectroscopy (XAS),[36,37] wavelength-modulated transmission spectroscopy,[38] photoluminescence spectroscopy,[39] electro-absorption measurement and absorption edge spectroscopy,[40] resonant ultraviolet photoelectron spectroscopy, and several other experimental techniques.[41-49] Amtout and Leonelli,[39] in their low-temperature absorption, time-integrated photoluminescence (PL), and resonant-Raman spectra have found the excitation energies of the intensity of the Raman scattering in the range of 2.737 to 3.031 eV. This anomalous excitation density dependence of the PL may be due in part to the direct dipole forbidden transition in Γ (conduction band) to Γ (valence band) to the more favorably allowed indirect transition. The experimentally measured direct band gaps range from 3.00 eV to 3.10 eV.

Although many theoretical calculations of the electronic properties of rutile $TiO_2$ have been reported in the literature,[1,3,7,8,31,50-70] there are, in general, obvious discrepancies between these previously calculated values and the corresponding experimental ones. The theoretically calculated, direct band gaps range from 1.67 to 3.25 eV (for LDA), 1.69 to 4.45 eV (for GGA), and over 3.4 to 13.05 eV (for HF). We are not aware of a previous report of a calculated, fundamental, indirect band gap for rutile $TiO_2$.

We aim to employ the Bagayoko – Zhao – Williams (BZW) method to calculate the band gap and other electronic properties of rutile $TiO_2$. The mathematical rigor of the method and the confirmation of our earlier successful predictions of band gaps and other properties of semiconductors[72-78] indicate that this work could shed light on the electronic properties of $TiO_2$.

Following this introductory and motivation in § 1, we describe our computational approach and the BZW method in § 2. In § 3, we present and discuss electronic energies and related properties of rutile $TiO_2$, as obtained by our self-consistent solution of the relevant *system of equation*s defining the local density approximation (LDA). We compare our findings to previous, corresponding theoretical and experimental ones. Finally, we summarize our results in § 4.

## 2. Computational Approach and the BZW Method

Our computational approach is characterized by our use of local density approximation (LDA) potential, the linear combination of atomic orbitals (LCAO) formalism, and the Bagayoko – Zhao – Williams (BZW)[73,74] method that solves, self-consistently, the system of equation defining LDA.

We employed the LDA potential of Ceperley and Alder[79] as parameterized by Vosko,Wilk, and Nusair.[80] We refer to it as the CA-VWN-LDA potential, using the first letters of the above authors. We should stress here that the use of an LDA potential is not sufficient if one wishes to obtain the correct eigenvalues of the Kohn-Sham equation: *Kohn and Sham*[81,82] *explicitly stated that the equations defining LDA have to be solved self-consistently.* They are (1) the equation giving the ground state charge density in terms of the wave functions of the occupied states, (2) the expression of the exchange correlation energy ($E_{xc}$) in the local density approximation (in terms of the ground state density), (3) the equation giving the exchange correlation potential ($V_{xc}$) as a functional derivative of $E_{xc}$ with respect to the ground state density, and (4) the Kohn-Sham equation. Once $E_{xc}$ is known, so is $V_{xc}$. Hence, after selecting an LDA potential (i.e., CA-VWN in our case here), the above system of four equations is reduced to one of two equations which are the equation giving the ground state density in terms of the wave functions of the occupied states and the Kohn-Sham equation. As noted below, this is the system we solve self-consistently with the BZW method.

In the linear combination of atomic orbitals (LCAO) approach, an unknown wave function for the solid state calculation is written as a linear combination of atomic orbitals. The radial parts of these orbitals are generally exponential or Gaussian functions resulting from self-consistent calculations of energy levels of the atomic or ionic species that are present in the solid under study. We use Gaussian functions and refer to that rendition of LCAO as the linear combination of Gaussian orbitals (LCGO). Many other calculations utilize the LCAO formalism, including those that employ plane waves.

The key difference between our computational approach and several others in the literature stems from our use of the BZW method to solve the applicable system of equations. As per our previous results and the ones discussed here for $TiO_2$, the agreements between our findings and experiments are due to the fact that the BZW method[74] adheres to the intrinsic, ground state nature of DFT, in general, and of LDA in particular, by searching for an optimal basis set that is *verifiably complete* for the description of the ground state. Other approaches employ a single trial basis set in their implementation of the LCAO formalism.

In the BZW method, we begin the solid state calculations with the minimum basis set, one that is just large enough to account for all the electrons in the system under study. Our self-consistent calculation with this basis set is followed by another where the basis set is augmented with one additional orbital from the atomic calculations. Taking the spin and angular symmetry into account, a radial orbital leads to 2, 6, 10, or 14 additional functions for s, p, d, and f states, respectively. The comparison of the occupied energies from Calculations I and II generally shows that they are different. A third, self-consistent calculation is performed with a basis set that includes that for Calculation II plus another orbital from the atomic calculations. This process of augmenting the basis set and of carrying out self-consistent calculations continues until a calculation, say N, is found to have the same occupied energies, within computational uncertainty of 50 meV, as calculation (N+1) that follows. This convergence of the occupied energies identifies the basis set of Calculation N as the optimal one. The optimal basis set is the smallest one with which all the occupied energies verifiably reach their respective minima. Put differently, the optimal basis set is the verifiably complete basis set that is converged with respect to the description of the occupied states.

In calculations with basis set larger than the optimal one, the ground state charge density does not change, nor do the Hamiltonian and the eigenvalues of the occupied states. Consequently, these calculations do not lower any occupied energies (as compared to the results obtained with the optimal basis set), even though they generally lead to some lower, unoccupied energies by virtue of the Rayleigh theorem.[73,77-78] This rigorous theorem states that when an eigenvalue equation is solved with two basis sets I and II, with set II larger than I and where I is entirely included in II, then the eigenvalues obtained with set II are lower or equal to their corresponding ones obtained with basis set I. This theorem explains the reasons that some unoccupied energies are lowered when the Kohn-Sham equation is solved with basis sets larger than the optimal one. Such a lowering of unoccupied energies with basis sets larger than the optimal one is fundamentally different from the one that occurs before the size of the basis set reaches that of the optimal one. The latter lowering is ascribed, at least in part, to the Hamiltonian, given that both the charge density and the Hamiltonian change, from one calculation to the next, before one reaches the optimal basis set while the former one is not.

Given that the DFT is a ground state theory constrained on the wave functions of the occupied states, seeking for the convergence of unoccupied states is noteworthy, meaningless (a converged excited states is achievable only within schemes that entirely go beyond the DFT). Reason for this is simple; there is no basis set that guarantees that the unoccupied states will converge simultaneously with the occupied ones. This is only possible for infinitely large basis set (say N to infinity) and can only be achieved within schemes that entirely go beyond the DFT. Of course, whether the excited states are converged or not, they are not meaningful.[83]

In fact, the BZW method essentially solves the *system of equations* describing LDA, as explicitly recommended by Kohn and Sham.[81] The BZW method solves self-consistently the system of equations, with the iterations for the Kohn-Sham equation embedded in those of the charge density equation. We recently found that narrow, upper valence band widths of wurtzite ZnO (around 3.5 eV or less), as reported by some single trial basis set calculations, are due to basis sets that are not complete for the description of the ground state, even though some of these basis sets are very large. The BZW method has been described in detail in the literature and employed in electronic property calculations of many semiconductors.[72-78]

With the above description of our approach, the following computational details permit the replication of our work. Our ab-initio, self-consistent calculations are non-relativistic. We utilized the electronic structure calculation package developed at the Ames Laboratory of the US Department of Energy (DOE), in Ames, Iowa.[84]

Rutile TiO$_2$ has a tetragonal structure (space group $D_{4h}^{14} - P4_2/mnm$ with Patterson symmetry $P4/mmm$) containing two titanium (cations) and four oxygen (anions) atoms, with the positions as indicated between parentheses: Ti: $(0,0,0)$; $(0.5,0.5,0.5)$ and O: $(0.3053, 0.3053, 0)$; $(-0.3053, -0.3053, 0)$; $(0.8053, 0.1947, 0.5)$; $(0.1947, 0.8053, 0.5)$ [85] (cf. Fig. 1). In the primitive unit cell (cf. Fig. 1), each Ti atom is surrounded by a slightly distorted octahedron of O atoms. The octahedra (TiO$_6$ which is the basic structural unit) centered respectively at $(0,0,0)$ and $(0.5,0.5,0.5)$ differ in orientation by a $90^0$ rotation about the c axis with the oxygen atoms forming a hexagonal closed-packed sublattice with half the octahedral sites being filled with Ti atoms. The titanium and oxygen atoms occupy the Wyckoff positions 2(*a*) and 4(*f*).[85-86]

Preliminary calculations indicated that in the solid, titanium is closer to Ti$^{2+}$ than to the neutral Ti. Similarly, oxygen species are O$^{1-}$ as opposed to the neutral O. The ionic nature of these species in TiO$_2$ could be partly inferred from the two column separation, in the periodic table, between Ti and O, on the one hand, and the relatively unshielded nuclei of O as compared to those of Ti, on the other hand. We first performed self-consistent calculations of the electronic properties and related functions for Ti$^{2+}$ and O$^{1-}$. Atomic orbitals utilized in these calculations, for the valence states, are given between parentheses: Ti$^{2+}$ (3s3p3d4s4p5s) and $O^{1-}$ (2s2p3s). Other atomic states with higher binding energies were treated as deep core states. In the basis sets for the valence states, (4p, 5s) and (3s) are unoccupied for $Ti^{2+}$ and $O^{1-}$, respectively. Nevertheless, these orbitals are included in the self-consistent LCAO calculations to allow for a reorganization of electronic cloud in the solid environment, including polarization. Our self-consistent calculations were performed at the room temperature experimental lattice parameters of a = 4.59373 Å and c = 2.95812 Å, with u = 0.3053.[86-87]

The self-consistent, ionic calculations led to trial ionic potentials for $Ti^{2+}$ and $O^{1-}$, respectively. These potentials were used to construct the input potential for rutile $TiO_2$. We used 16 Gaussian functions for the s and p states and 14 for the d states for $Ti^{2+}$ and utilized 17 Gaussian functions for the s and p states for $O^{1-}$. A mesh of 60 *k* points, with proper weights in the irreducible Brillouin zone, was employed in the self-consistent (solid calculations) iterations. In total, 141 weighted k-points were used in the band structure calculations, and a total of 147 weighted k-points were employed to generate the energy eigenvalues for the electronic density of states computations using the linear, analytical tetrahedron method.[88] The k-points were chosen along the high symmetry points in the Brillouin zone. We also calculated the partial density of states using the Mulliken partitioning method.[89] The self-consistent potentials converged to a difference around $10^{-5}$ after about 60 iterations.

We carried out structural optimization for $TiO_2$. In calculating the lattice parameters, we utilized the Murnaghan's equation of state.[90-91]

### 3. Results and Discussions

Following the BZW method, we performed successive, self-consistent calculations of the electronic properties of $TiO_2$. We performed a total of six calculations, beginning with the one employing the minimum basis set. The optimal basis set was that of Calculation IV. We recall that occupied energies reach their minimum when the optimal basis set is used, in case they had not already done so. The electronic energy bands and related properties discussed below are as obtained in Calculation IV. Figures 2, 3, 4, and 5, respectively show the electronic energies, the total density of sates, the partial densities of states, and the electron density in the (100) plane.

From Fig. 2, it can be seen that the minimum of the conduction band occurs at the R point while the maximum of the valence band is at the Γ point, resulting in a predicted, fundamental, indirect gap of 2.95 eV. However, our calculated direct band gap of 3.05 eV at the Γ point is only larger by 0.10 eV. These results somewhat corroborate reports of direct and indirect transitions that are nearly degenerate.[38,47,92-93] Our calculated, direct gap of 3.05 (i.e., 3.046 eV) is in agreement with experiment. Specifically, over fourteen experimental works report gaps ranging from 3.0 to 3.10 eV, as shown in Table I. Table I provides a comparison of our findings to other theoretical results which mostly underestimate the band gap. Table II shows our calculated band widths at the Γ and the above band gaps along with results from some previous theoretical and experimental reports.

We recall that the electronic structure in Fig. 2 was obtained by using the room temperature experimental lattice constants ($a_o$ and $c_o$) given above. We further examined whether or not the position of the shallow minimum of the conduction band is strongly dependent on the lattice constants. We performed two calculations with 1% increase and 2% decrease in both lattice constants from their room temperature values of $a_0$ and $c_0$. In both cases, the minimum remained at the R point. The indirect band gap decreased to 2.73 eV following the 1% increase and increased to 3.02 eV after the 2% contraction.

A distinctive feature of the electronic band structure consists of groups of bands that are well separated; this feature is apparent in Figs. 2 and 3. The lowest laying valence bands are mostly of O – 2s character with a little hybridization with the Ti – p and Ti – s states, respectively; the partial densities of states in Fig. 4 show this feature. As per the content of Fig. 4, the upper valence bands emanate from a very strong hybridization between O – 2p and Ti-3d states. The group of lowest conduction bands is primarily of O – 2p and Ti – 3d states. These observations suggest that the excitation across the band gap involves both O-2p and Ti-3d states, in agreement with earlier findings of Mo and Ching.[3] The electronic band structure reveals the conduction band minimum at the Γ point consisting of two energetically close bands. The calculated energy difference between these two bands is only 0.12 eV, in basic agreement with the 0.11 eV observed by Persson and da Silva.[60]

Figure 5 shows the electron density of $TiO_2$ in the (100) plane. This plot merely illustrates the possibilities for further exploration of the electronic properties of $TiO_2$ with plots of this type in different high symmetry planes.

In Fig. 6, we show the calculated total energy as a function of the ratio *c/a*, obtained using volume-constrained total energy minimization. Our calculated ratio of $c_o/a_o$ at the equilibrium volume of 32.1605 ($Å^3/TiO_2$) is 0.6381 (0.64). This latter value is exactly the same as the experimental one of 0.64 for $c_o/a_o$. Our calculated equilibrium lattice constants are 4.6538 Å and 2.9697 Å for $a_o$ and $c_o$, respectively. The calculated, internal parameter *u* is 0.3051, basically the same as the experimental value of 0.3053.[86] The calculated values of the lattice parameters are within the ranges of experimentally reported room temperature lattice parameters of rutile $TiO_2$ which range from 4.588 to 4.657 Å and 2.95407 to 2.967 Å for *a* and *c*, respectively.[86] Using the Murnghan equation of state, we calculated a bulk modulus of 214.97 GPa. This result is basically the same as the reported, experimental value of 216.[95] Our calculated pressure derivative ($B`_o$) is 4.38. This value is, however, much lower than the experimental value of 6.84, as obtained by single-crystal ultrasonic experiments.[96] This large discrepancy, we suggest, stems from two possible difficulties. The first one is related to the visible flatness, around the minimum total energy, of the total energy versus c/a curve; it directly results in large fitting uncertainties. The second difficulty is that of obtaining accurate, experimental values for the $B'_o$.

The electron effective mass of rutile tatania has not been unambiguously determined. Our calculated electron effective masses in the Γ-M, Γ–Z, and Γ-X directions are in the range: 1.13 to 1.20, 0.62 to 0.64, and 1.14 to 1.20 $m_0$, respectively. The apparent anisotropy in the effective mass of $TiO_2$ is expected in a tetragonal structure.

Table II compares our numerical results for the electronic properties of rutile $TiO_2$ electronic structure with those from experiment and other theoretical calculations. Our calculated results compare more favorably with experiment. We present, in Table III, the calculated eigenvalues at the high symmetry points in the Brillouin zone. They are expected to enable comparisons of our results with future theoretical and experimental ones.

**4.    Conclusions**

We performed a first principle computational study of the electronic and related properties of rutile $TiO_2$ within density functional theory (DFT), using a local density approximation potential. We utilized the linear combination of atomic orbitals (LCAO) formalism. Our use of the BZW method led to an optimal basis set that is verifiably complete for the description of the ground state. In addition to the electronic band structure, we obtained the electron effective mass and the total and partial densities of states.

Our ab-initio, self-consistent LDA-BZW calculations led to ground state electronic and related properties that mostly agree with experiment. Specifically, we found that the fundamental band gap of rutile $TiO_2$ is an indirect band gap of 2.95 eV, from Γ to R. The calculated, direct band gap of 3.05 eV, at the Γ point, is in excellent agreement with concordant findings of over 14 experiments – as shown in Table I. Our structural optimization of $TiO_2$ reproduced values that are in perfect agreement with experiment. We expect the detailed energies at the high symmetry points, in Table III, to enable future comparisons with experimental measurements, from optical absorption to X - ray studies of the semi-core states.

**Acknowledgements**

Discussions with J. P. Perdew have been beneficial. This work was funded in part by the Louisiana Optical Network Initiative (LONI, Award No. 2-10915), the Department of the Navy, Office of Naval Research (ONR, Award Nos. N00014-08-1-0785 and N00014-04-1-0587), the National Science Foundation [Award Nos. 0754821, EPS-1003897, and NSF (2010-15)-RII-SUBR], and Ebonyi State, Federal Republic of Nigeria (Award No: EBSG/SSB/FSA/040/VOL. VIII/039).

Table I: Comparison of our calculated LDA-BZW band gap with other theoretical and room temperature experimental band gaps of rutile $TiO_2$ at the Γ point. Unless otherwise stated, all band gaps are direct. Reference numbers are indicated as superscripts in column 1.

| Authors | Band Gap $E_g$ (eV) | Method | Potential |
|---|---|---|---|
| **Theoretical Results** | | | |
| **This Work** | 2.95 (Indirect); 3.05 (Direct) | LCAO-BZW | LDA-DFT |
| Fox et al. [2010][8] | 2.46 | SCC-DFTB | Semi-Empirical |
| Glassford and Chelikowsky [1992][51] | 2.00 | PW-PP | LDA-DFT |
| Labat et al. [2007][1] | 1.88, 1.83, 2.14 | PBE-LCAO | GGA-DFT |
| | 12.14, 12.21, 13.05 | HF-LCAO | HF |
| | 4.05, 4.02, 4.45 | PBE0-LCAO | GGA-DFT |
| | 1.85, 1.82, 2.12 | LCAO | LDA-DFT |
| | 3.53, 3.50, 3.92 | B3LYP | GGA-DFT |
| | 1.67 | PAW | LDA-DFT |
| | 1.69 | PBE-PAW | GGA-DFT |
| Mo and Ching [1995][3] | 1.78 | SC-OLCAO | LDA-DFT |
| Vogtenhuber et al. [1994][50] | 1.99 | FLAPW | LDA-DFT |
| Silvi et al. [1991][52] | >3.40 | HF-PP | HF |
| Poumellec et al. [1991][53] | 2.0 | LMTO | ASA |
| Paxton and Thien-Nga [1998][54] | 1.80 | FPLMTO | LSDA |
| Islam et al. [2007][7] | 3.54 | DFT-HF Hybrid | PW1PW |
| | 1.90 | PWGGA | GGA |
| Cho et al. [2006][55] | 1.70 | PP-PAW | LDA-DFT |
| Lee et al. [1994][56] | 1.87 | Variational Density-Functional Perturbation | LDA-DFT |
| Grunes et al. [1982][31] | 2.80 | PP | Tight Band |
| Kasowski and Tait [1979][57] | 3.25 | LCMTO | LDA-DFT |
| Mattioli et al. [2008][58] | 2.00 | PW-PBE | LSD-GGA+U |
| Shirley et al. [2010][59] | 1.86 | PP-PW | GGA-PBE |
| Persson and da Silva [2005][60] | 1.80 | FPLAPW | LDA-DFT |
| | 2.97 | FPLAPW | LDA+$U^{SIC}$ |
| Kesong et al. [2007][62] | 1.85 | PP | GGA-DFT |
| Shao [2008][63] | 1.87 | PP-PW | PBE-GGA |
| | 2.03 | PP-PW | PBE-WC-GGA |
| **Experimental Results (Band Gaps are Measured at Room Temperature)** | | | |
| Cronemeyer [1952][42] | 3.05 | Electrical and Optical methods | N/A |
| Tang et al. [1977][26] | 3.06 | XPS | N/A |
| Persson and da Silva [2005][60] | 3.08 | dc Magnetron Sputtering And Sol-Gel Technique | N/A |
| Pascual et al. [1978][40] | 3.062 | High Resolution Absorption Edge Spectra | N/A |
| Tang et al. [1995][43] | 3.03 | Polarized Optical | N/A |

| | | Transmission | |
|---|---|---|---|
| Lu et al. [1995][44] | 3.10 | Adsorption Photodesorption of Oxygen | N/A |
| Rocker et al. [1984][28] | 3.00 | EELS | N/A |
| Knotek and Feibelman [1978][45] | 3.00 | Ion Desorption | N/A |
| Fischer [1972][46] | 3.03 | X - ray Emission and Absorption Band Spectra | N/A |
| Tsutsumi et al. [1977][34] | ~3.00 | Emission and Absorption Spectra | N/A |
| Pascual et al. [1977][47] | 3.031 | Absorption Spectra | N/A |
| Arntz and Yacoby [1966][48] | 3.00 | Electroabsorption Measurement | N/A |
| Amtout and Leonelli [1995][39] | 3.031 | Photoluminescence, and Resonant-Raman-Scattering Spectra | N/A |
| Burdett et al. [1987][49] | 3.00 | Pulsed Neutron Diffraction | N/A |
| Tait and Kasowski [1979][32] | 3.00 | UPS, LEED and AES | N/A |
| Kowalczyk et al. [1977][25] | 3.06 | XPS | N/A |

**Table II**: Comparisons of some important parameters of the electronic structure of bulk rutile $TiO_2$

| Property (eV) | LDA-BZW | GGA-DFT[a] | LDA-DFT | Experiment |
|---|---|---|---|---|
| Upper valence bandwidth | 5.04 | 5.69 | 5.70[b]; 6.22[c] | 5-6[d], 5.4[f] |
| Lower valence bandwidth | 1.95 | 1.79 | 1.80[b]; 1.94[c] | 1.90[d] |
| Lower conduction bandwidth | 5.30 | N/A | N/A | N/A |
| Total width of the valence band at Γ | 17.71 | 18.13 | 17.00[b]; 17.98[c] | 16-18[e] |
| Direct Band gap at Γ-Γ | 3.05 | 1.88 | 2.00[b]; 1.78[c] | 3.06[d] |
| Indirect, Fundamental Band gap (Γ-R) | 2.95 | N/A | N/A | N/A |

[a]Reference 1; [b]Reference 51; [c]Reference 3; [d]Reference 13,25,46; [e]Reference 45; [f]Reference 40

**Table III:** Calculated eigenvalues (in eV) at the high symmetry points, for rutile $TiO_2$. The eigenvalues are obtained by setting the Fermi energy, which occurred at Γ, equal to zero.

| Γ | X | R | Z | M | A |
|---|---|---|---|---|---|
| -17.706 | -16.901 | -16.708 | -16.319 | -16.989 | -16.440 |
| -16.492 | -16.901 | -16.708 | -16.319 | -16.989 | -16.440 |
| -15.760 | -16.050 | -15.838 | -16.265 | -15.924 | -16.122 |
| -15.760 | -16.050 | -15.838 | -16.265 | -15.924 | -16.122 |
| -5.045 | -4.316 | -4.487 | -3.883 | -4.952 | -4.314 |
| -4.887 | -4.316 | -4.487 | -3.883 | -4.952 | -4.314 |
| -4.874 | -3.527 | -3.684 | -3.186 | -3.657 | -3.826 |
| -3.612 | -3.527 | -3.684 | -3.186 | -3.657 | -3.826 |
| -3.612 | -3.314 | -2.601 | -3.0257 | -2.857 | -3.012 |
| -2.243 | -3.314 | -2.601 | -3.0257 | -2.857 | -3.012 |
| -2.243 | -2.456 | -2.051 | -2.606 | -2.364 | -2.362 |
| -1.560 | -2.456 | -2.052 | -2.606 | -2.364 | -2.362 |
| -0.196 | -1.251 | -1.934 | -2.349 | -0.883 | -1.527 |
| -0.175 | -1.251 | -1.934 | -2.349 | -0.883 | -1.527 |
| -0.0002 | -0.861 | -0.924 | -1.319 | -0.515 | -0.419 |
| 0 | -0.861 | -0.924 | -1.319 | -0.515 | -0.419 |
| 3.046 | 3.693 | 2.950 | 3.978 | 3.147 | 3.545 |
| 3.167 | 3.693 | 2.950 | 3.978 | 3.147 | 3.545 |
| 3.167 | 3.850 | 3.989 | 4.016 | 3.180 | 4.046 |
| 3.233 | 3.850 | 3.989 | 4.016 | 3.180 | 4.046 |
| 3.480 | 4.381 | 4.679 | 4.675 | 4.680 | 4.103 |
| 4.965 | 4.381 | 4.679 | 4.675 | 4.680 | 4.103 |
| 5.329 | 6.604 | 5.379 | 6.024 | 6.268 | 6.388 |
| 7.078 | 6.604 | 5.379 | 6.024 | 6.268 | 6.388 |
| 7.078 | 6.788 | 7.686 | 6.551 | 7.0881 | 6.512 |
| 8.021 | 6.788 | 7.686 | 6.551 | 7.0881 | 6.512 |

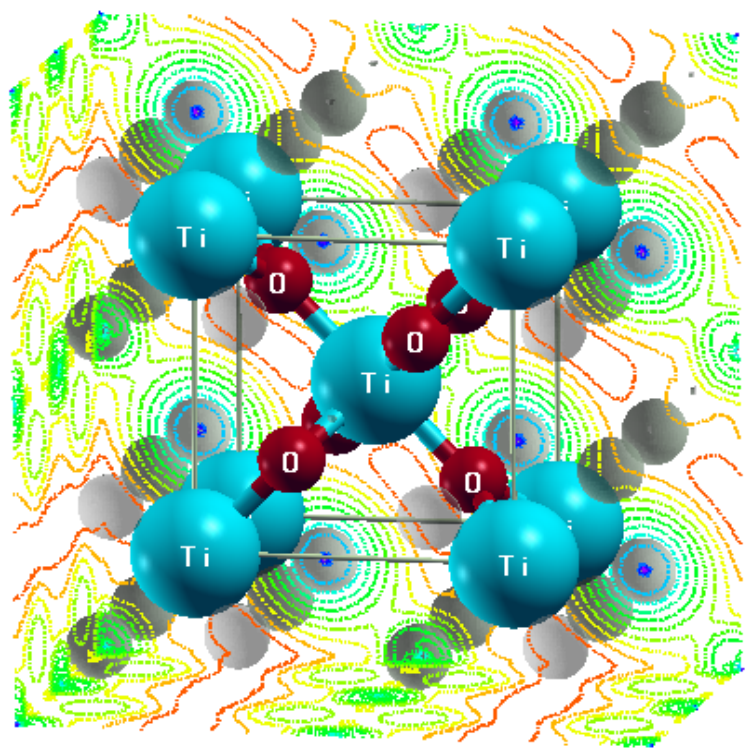

Fig. 1. (Color online) The tetragonal unit cell of TiO$_2$ with the iso-surface, at lattice parameters of a = 4.59373 Å and c = 2.95812 Å, with u = 0.3053. (Figure has been drawn using xcrysden (Ref. 94)).

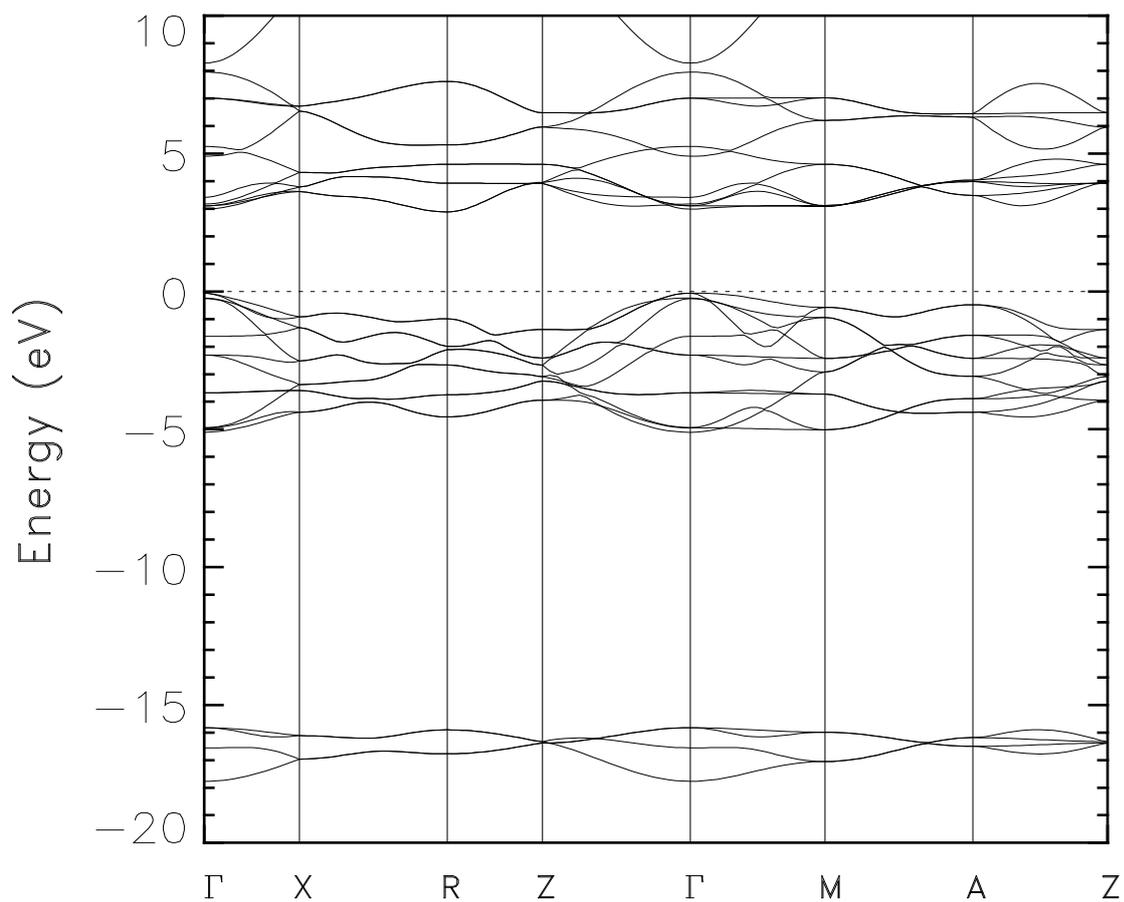

**Figure 2.** The calculated, electronic energy bands of rutile $TiO_2$ as obtained with the optimal basis set. The calculated direct band gap of 3.05 eV is practically the same as the experimental one. (i.e., 3.0 eV to 3.10 eV). The minimum gap, from $\Gamma$ to R, is 2.95 eV. The horizontal, dashed line indicates the position of the Fermi energy ($E_F$) which has been set equal to zero.

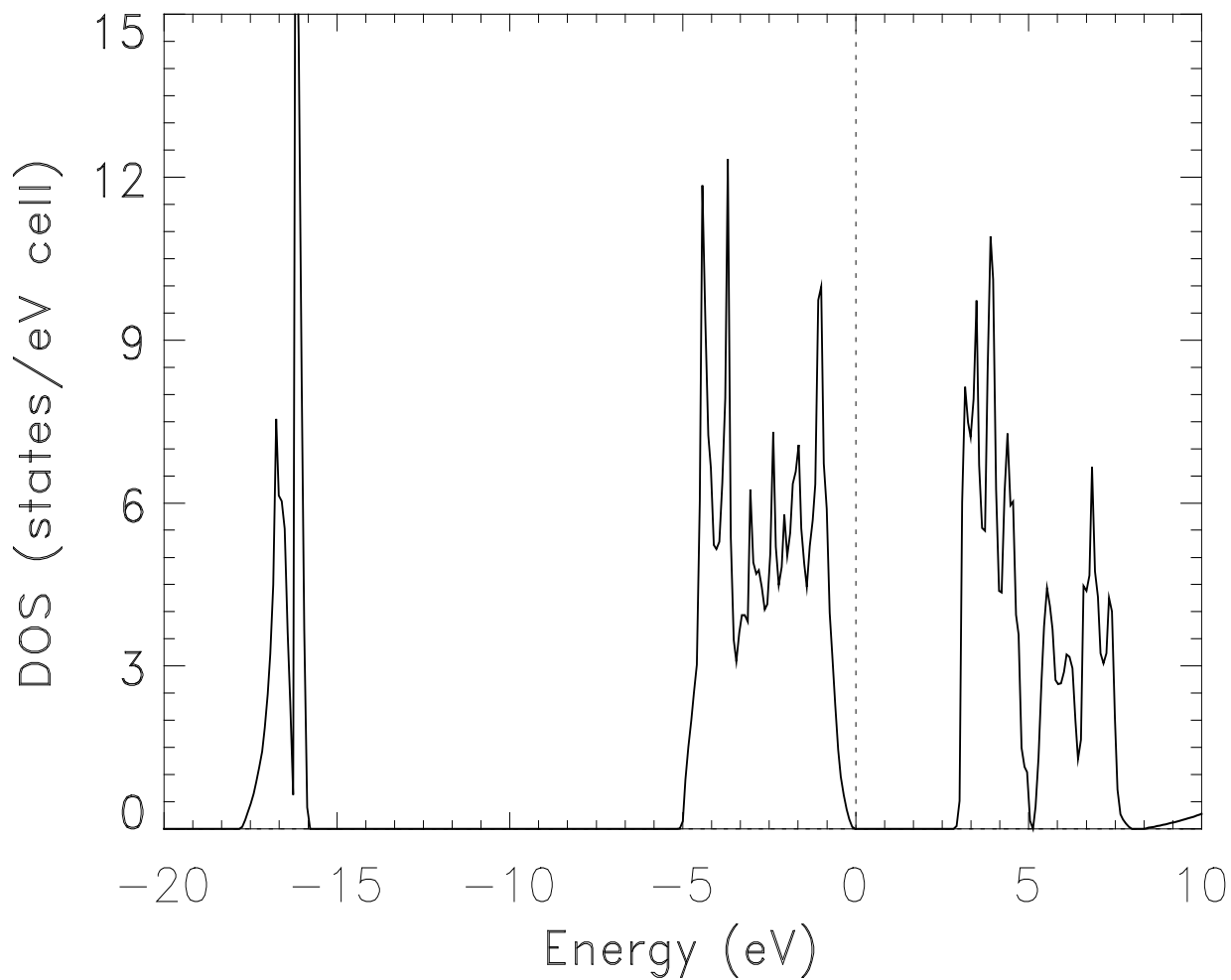

**Figure 3.** The total density of states (DOS) of rutile TiO$_2$, as obtained from the bands shown in Figure 2. The vertical, dashed line indicates the position of the Fermi energy (E$_F$) which has been set equal to zero.

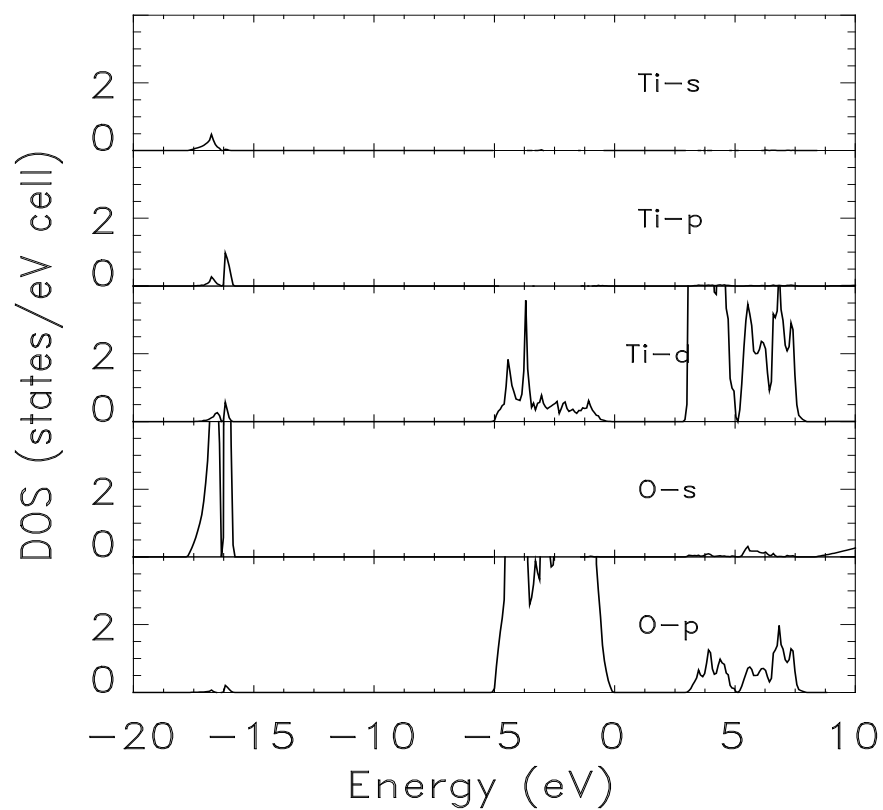

**Fig. 4**: The partial density of states (pDOS) of rutile $TiO_2$, as obtained from the bands shown in Figure 2. The position of zero eV indicates that of the Fermi energy ($E_F$) which has been set equal to zero.

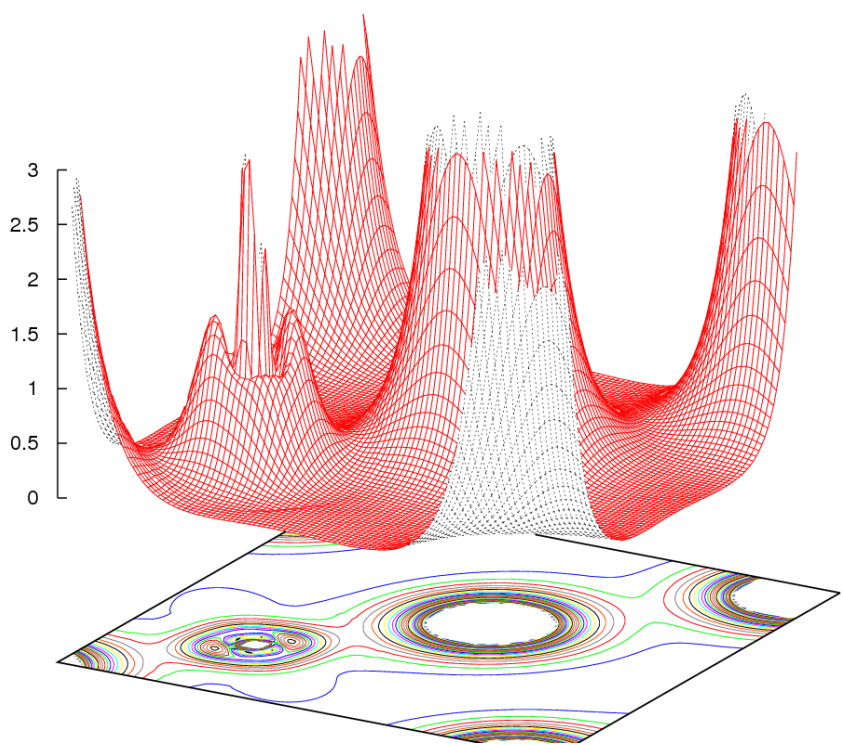

**Fig. 5**: (Color online) The electron density of $TiO_2$ along the (100) plane. (Figure has been drawn using xcrysden (Ref. 94)).

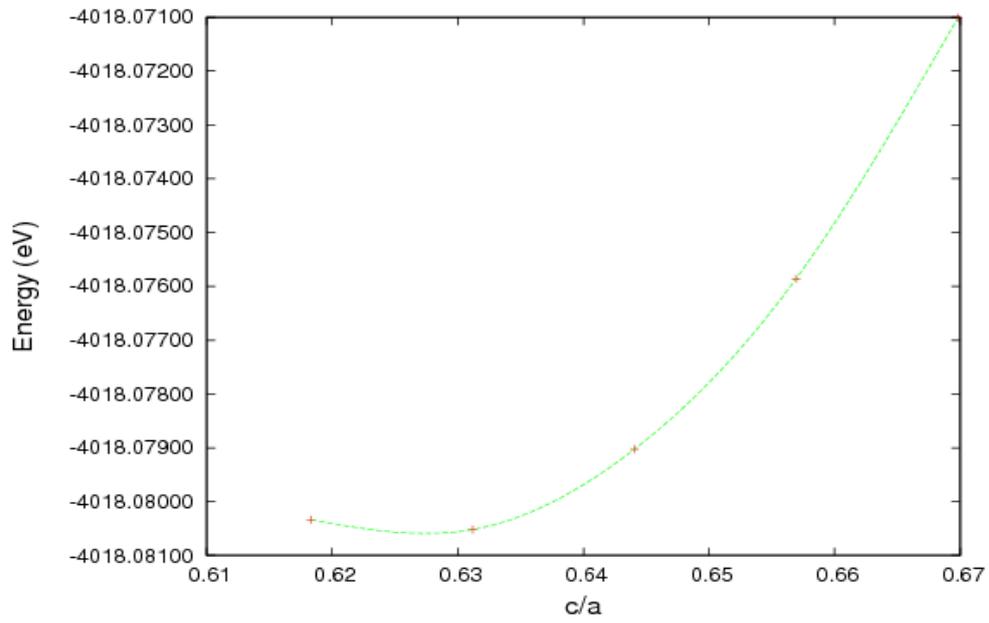

**Fig. 6**: (Color online) The calculated total energy as a function of the ratio of *c/a*.